\documentclass[3p,times,twocolumn]{elsarticle}

\usepackage{ecrc}

\volume{00}

\firstpage{1}

\journalname{Nuclear Physics B Proceedings Supplement}

\runauth{C. Hagedorn}

\jid{nuphbp}

\jnltitlelogo{Nuclear Physics B Proceedings Supplement}

\usepackage{amssymb}
\usepackage[figuresright]{rotating}

\begin{document}

\begin{frontmatter}



\dochead{}

\title{Charged lepton flavour violation in supersymmetric and holographic composite Higgs models with flavour symmetries}


\author[label1]{C. Hagedorn} 

\address[label1]{Dipartimento di Fisica e Astronomia `G.~Galilei', Universit\`a di Padova, Via Marzolo~8, I-35131 Padua, Italy and SISSA, Via Bonomea 265, I-34136 Trieste, Italy}

\begin{abstract}

We discuss charged lepton flavour violating processes such as $\mu\to e\gamma$ in supersymmetric extensions of the Standard Model and in models
with gauge-Higgs unification and a warped extra dimension. In both classes of models the processes turn out to be compatible with present experimental
bounds, but can be detected by future experiments, thanks to the presence of a flavour symmetry $G_f$ which constrains the form of the relevant couplings.
The symmetry $G_f$ is chosen to be finite, discrete and non-abelian and also helps to predict the peculiar lepton mixing pattern.

\end{abstract}


\end{frontmatter}

% \linenumbers

\section{Introduction}
\label{intro}

The existence of three generations of elementary particles,
 	the strong hierarchy among the charged fermion masses, the small mixing among quarks and the lepton mixing pattern with
	two large and one small mixing angle can be accommodated, but  not explained within the 
	Standard Model (SM) and many of its extensions. A possible explanation of these features
	are  flavour symmetries $G_f$ that act on the space of the three generations. Finite discrete non-abelian
	groups turned out to be prime candidates for the explanation of the peculiar lepton mixing pattern, especially
	if they are broken to different (non-trivial) subgroups $G_e$ and $G_\nu$ in the charged lepton and the neutrino
	sector, respectively. Since $G_e \neq G_\nu$, $G_f$ is fully broken at low energies.
	
	In theories beyond the SM additional degrees of freedom and/or additional interactions are present, which induce
	flavour violating processes, e.g. $\mu\to e\gamma$,  
	that are tightly
	constrained by experiments. Flavour symmetries $G_f$ can play a crucial role in constraining the couplings 
	relevant for such processes. In the following we discuss this aspect in two different classes of models:
	extensions of the Minimal Supersymmetric SM with the flavour symmetry $A_4$ \cite{FHLM,HMP,AFMS} and
	models with gauge-Higgs unification and a warped extra dimension  (which are realizations of
	holographic composite Higgs models) with $G_f=X \times Z_N$ and $X$ non-abelian \cite{HS1,HS2}. 

\section{Supersymmetric models with $G_f=A_4$}
\label{susy}

 Tri-bimaximal lepton mixing ($\sin^2 \theta_{12}=1/3$, $\sin^2 \theta_{23}=1/2$, $\theta_{13}=0$) can be explained with 
	the flavour symmetry $A_4$, if  left-handed (LH) leptons transform as ${\bf 3}$ under $A_4$ and the latter is broken to $Z_3$ in the charged lepton and to $Z_2$
	in the neutrino sector. 
	This symmetry breaking pattern can be achieved in a simple way in supersymmetric models, if
	the group $A_4$ is broken spontaneously through vacuum expectation values (VEVs) of gauge singlets which transform
	under $A_4$, so-called flavons. The parameter $\xi$ quantifies the size of the breaking of $A_4$
	(i.e. the generic VEV of the flavons over the cutoff scale $\Lambda$
	of the theory). For $\xi \sim 0.1$
	the reactor mixing angle is $\theta_{13} \sim 0.15$, in accordance with experimental data. Higher-dimensional operators with several flavons induce 
	corrections to the leading order (LO) results which are proportional to powers of $\xi$.
	Right-handed (RH) charged leptons are in different singlets of $A_4$
	and are charged under a Froggatt-Nielsen symmetry $U(1)_{FN}$ so that the charged lepton mass hierarchy is achieved.

	We consider a theory in which the symmetry $A_4 \times U(1)_{FN}$ also determines the flavour structure of the soft supersymmetry breaking terms, soft masses
	and A-terms \cite{FHLM,AFMS}. The soft masses $m_{\tilde L}^2$ for LH sleptons are flavour universal at LO, since they form a
	triplet ${\bf 3}$ under $A_4$, whereas soft masses $m_{\tilde R}^2$ for RH charged sleptons are flavour diagonal.  
	A-terms are diagonal in flavour space and $A_{\alpha\alpha} \propto m_\alpha$ for $\alpha=e, \mu, \tau$.
	Corrections from higher-dimensional operators lead to deviations
	from these simple structures so that the soft masses $m_{\tilde L}^2$ acquire off-diagonal elements proportional to $\xi^2$, while $(m_{\tilde R}^2)_{\alpha\beta} \propto \xi \times m_\alpha/m_\beta$
	with $m_\alpha < m_\beta$ in the physical (lepton mass) basis.
	 The A-terms are non-diagonal as well and $A_{\alpha\beta} \propto  \xi \times m_\alpha$ for $m_\alpha < m_\beta$, while $A_{\alpha\beta}$ for $m_\alpha>m_\beta$ is either proportional to $\xi^2 \times m_\alpha$ or  
	to $\xi \times m_\alpha$.\footnote{The latter type of contribution, however, can vanish under certain assumptions, made about the flavon potential.}  
	
	 These non-vanishing off-diagonal elements in the soft masses and A-terms induce flavour violating processes, e.g. $\mu\to e\gamma$
       can arise from  $(m_{\tilde L}^2)_{\mu e} \neq 0$. An analytic estimate shows that the branching ratio $\mathrm{BR} (\mu\to e\gamma) \propto \xi^2 \; \mbox{or} \; \xi^4$
       depending on the existence of the above-mentioned contribution to the A-terms. $\mathrm{BR}(\tau\to (e, \mu) \gamma)$ should be of the same order of magnitude as $\mathrm{BR} (\mu\to e\gamma)$ \cite{FHLM,AFMS} and thus
       are in general too small for detection in the near future.
	 In our numerical analysis \cite{FHLM,AFMS} we have computed $\mathrm{BR} (\mu\to e\gamma)$, scanning over a large range of soft scalar and gaugino masses $(m_0, M_{1/2})$
	 and setting $\xi\approx 0.076$ and $\tan\beta=2$ or $\tan\beta=15$.  
	For $\tan\beta=2$ and $m_0=200$ GeV $M_{1/2}$ has to be larger than $500$ GeV, while for $m_0=5000$ GeV smaller values of $M_{1/2}$ are sufficient for passing the 
 current experimental limit from MEG. For larger $\tan\beta$ $\mathrm{BR}(\mu\to e\gamma)$ increases approximately like $\tan^2\beta$.
Including the renormalization group running effects from (three) RH neutrinos can increase $\mathrm{BR} (\mu\to e\gamma)$ by a factor $10^{1 \div 2}$ \cite{HMP,AFMS}.

\section{Models with gauge-Higgs unification, a warped extra dimension and $G_f=X \times Z_N$}
\label{hchm}

The gauge group in the bulk is $SO(5) \times U(1)_X$ (without $SU(3)_c$), necessary 
       for gauge-Higgs unification, and it is broken at the ultraviolet (UV, Planck) and infrared (IR, TeV) boundaries to $SU(2)_L \times U(1)_Y$
       and $SO(4) \times U(1)_X \times P_{LR}$, respectively. The bulk flavour symmetry $G_f=X \times Z_N$, $N \geq 3$, is broken at the boundaries
       to $Z_2 \times Z_2 \times Z_N$ and to $Z_N^{(D)}$ which is the diagonal subgroup of
       $Z_N \subset X$ and the additional $Z_N$.  Auxiliary (flavour universal and abelian) symmetries $G_a$, $G_{a,\mathrm{UV}}$ and $G_{a,\mathrm{IR}}$
       reduce the number of couplings \cite{HS1,HS2}.
 
  \begin{figure}
  \begin{center}
\includegraphics[width=3in]{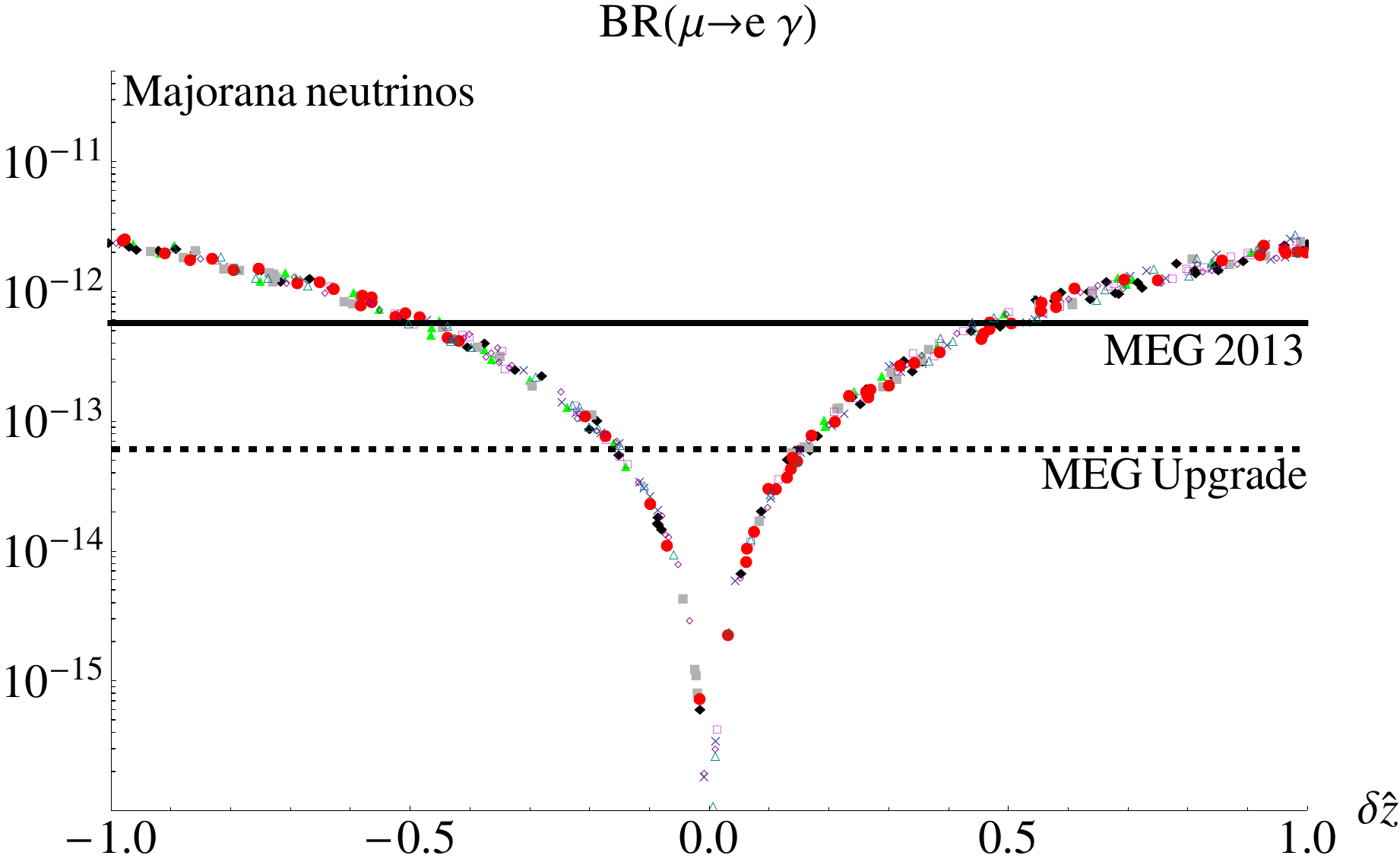}\\[0.15in]
  \includegraphics[width=3in]{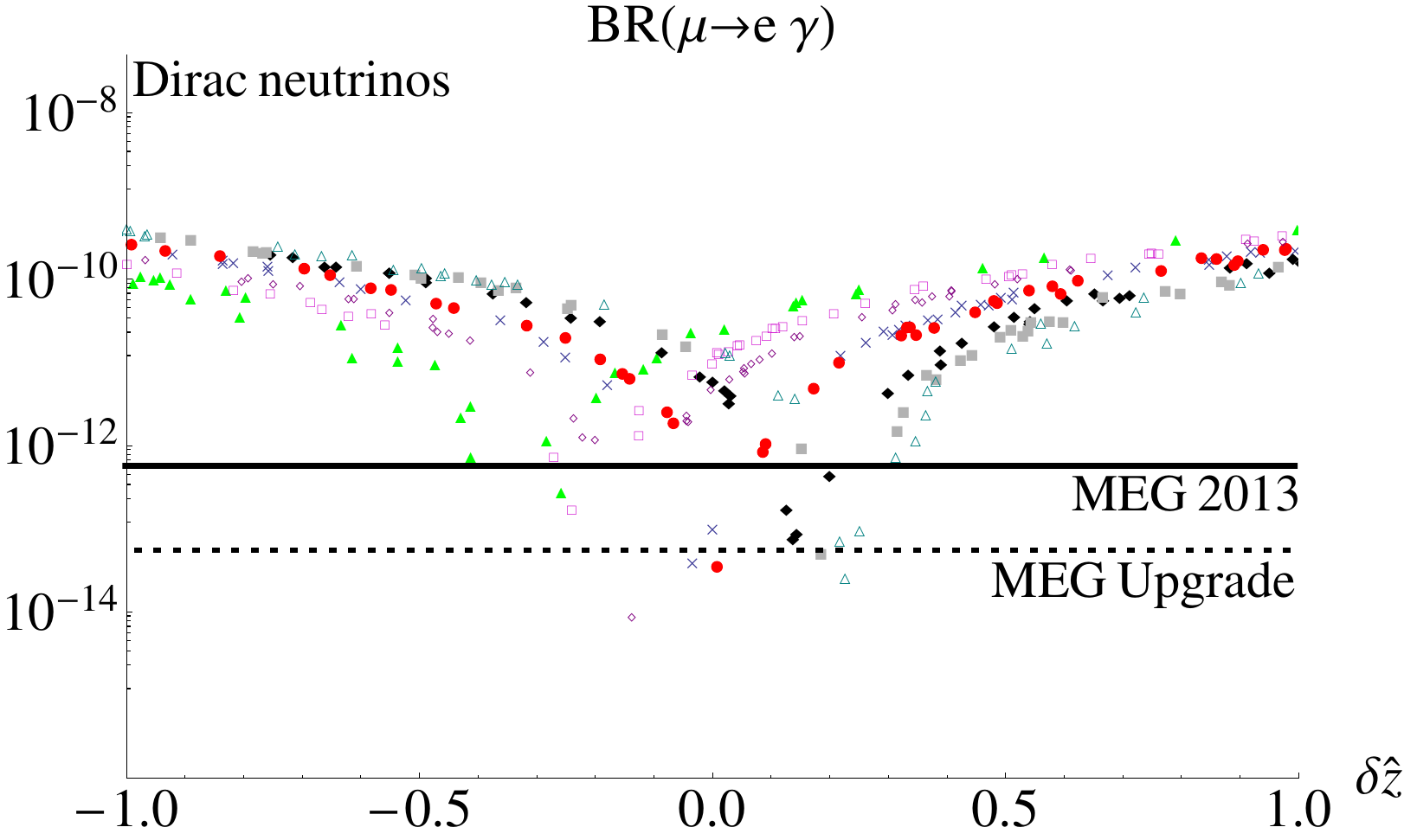}
  \end{center}
\caption[]{Results for $\mathrm{BR} (\mu\to e\gamma)$ for Majorana (upper) as well as Dirac neutrinos (lower figure) depending on the size $\delta \hat z$ of the boundary kinetic terms. 
The different symbols stand for different flavour symmetries $X$ (and mixing patterns ${\mathrm M} \, \mathrm{i}$): red dots for $X=S_4$,
blue crosses for $X=A_5$, filled (open) light (dark) green triangles for $X=\Delta (96)$ and pattern ${\mathrm M}1$ (${\mathrm M}2$), filled (open) black (violet) diamonds for $X=\Delta (384)$ and pattern ${\mathrm M}3$ (${\mathrm M}4$)
and filled (open) grey (pink) squares for $X=\Delta (1536)$ and pattern ${\mathrm M}5$ (${\mathrm M}6$). Current and expected experimental limits from MEG 2013 and MEG Upgrade are shown.}
\label{fig3}
\end{figure}
       
     The particles of the SM are identified with the zero modes of the $SO(5)$ multiplets $\xi_{\nu,\alpha}\sim {\bf 1}$, $\xi_{L,\alpha}\sim {\bf 5}$ and $\xi_{e,\alpha}\sim {\bf 10}$ present in the bulk [4,5] (all are neutral under $U(1)_X$). In particular,
    RH neutrinos reside in the singlets $\xi_{\nu,\alpha}$, LH lepton doublets in $\xi_{L,\alpha}$ with $T_{3R}=-1/2$ and RH charged leptons in  $\xi_{e,\alpha}\sim {\bf 10}$ with $T_{3R}=-1$. The multiplets  $\xi_{\nu,\alpha}$
    and $\xi_{L,\alpha}$ transform as ${\bf 3}$ under $X$ and trivially under $Z_N$, while $\xi_{e,\alpha}$ are singlets under $X$ and distinguished with the help of $Z_N$. In this way, lepton mixing turns out to be determined
    by the breaking of $G_f$ to $Z_2 \times Z_2 \times Z_N$ and $Z_N^{(D)}$ at the boundaries and the charged lepton mass hierarchy is achieved via the appropriate localization of $\xi_{e,\alpha}$ in the extra
    dimension. 
    
One can construct models for both types of neutrinos, Majorana and Dirac \cite{HS1,HS2}. The crucial difference lies in the assignment of the boundary conditions of the singlet state with $T_{3L}=T_{3R}=0$ contained in  $\xi_{L,\alpha}$,
i.e. $(-+)$ for Majorana and $(+-)$ for Dirac neutrinos. 
Charged lepton masses remain unaffected by this choice of possibilities and arise through mass terms localized at the IR boundary. In the model for Majorana neutrinos an accidental $Z_2$ symmetry
     at the IR boundary which exchanges multiplets with the same transformation properties under $SO(4)$ further reduces the number of parameters.
 
 We have considered the flavour symmetries $X=S_4, \, \Delta(96), \, \Delta(384), \, \Delta(1536)$ with $N=3$ and $X=A_5$ with $N=5$. Especially for $X=\Delta(1536)$ lepton mixing angles are
   very close to the experimental data at LO: $\sin^2 \theta_{12} \approx 0.342$, $\sin^2 \theta_{23} \approx \left\{ 0.387 \, (\mathrm{M}5) \, , \, 0.613 \,  (\mathrm{M}6) \right\}$, $\sin^2 \theta_{13} \approx 0.0254$.
   Including the effect of Kaluza-Klein states these results change only slightly in both models.

\begin{figure}
\begin{center}
   \includegraphics[width=2.9in]{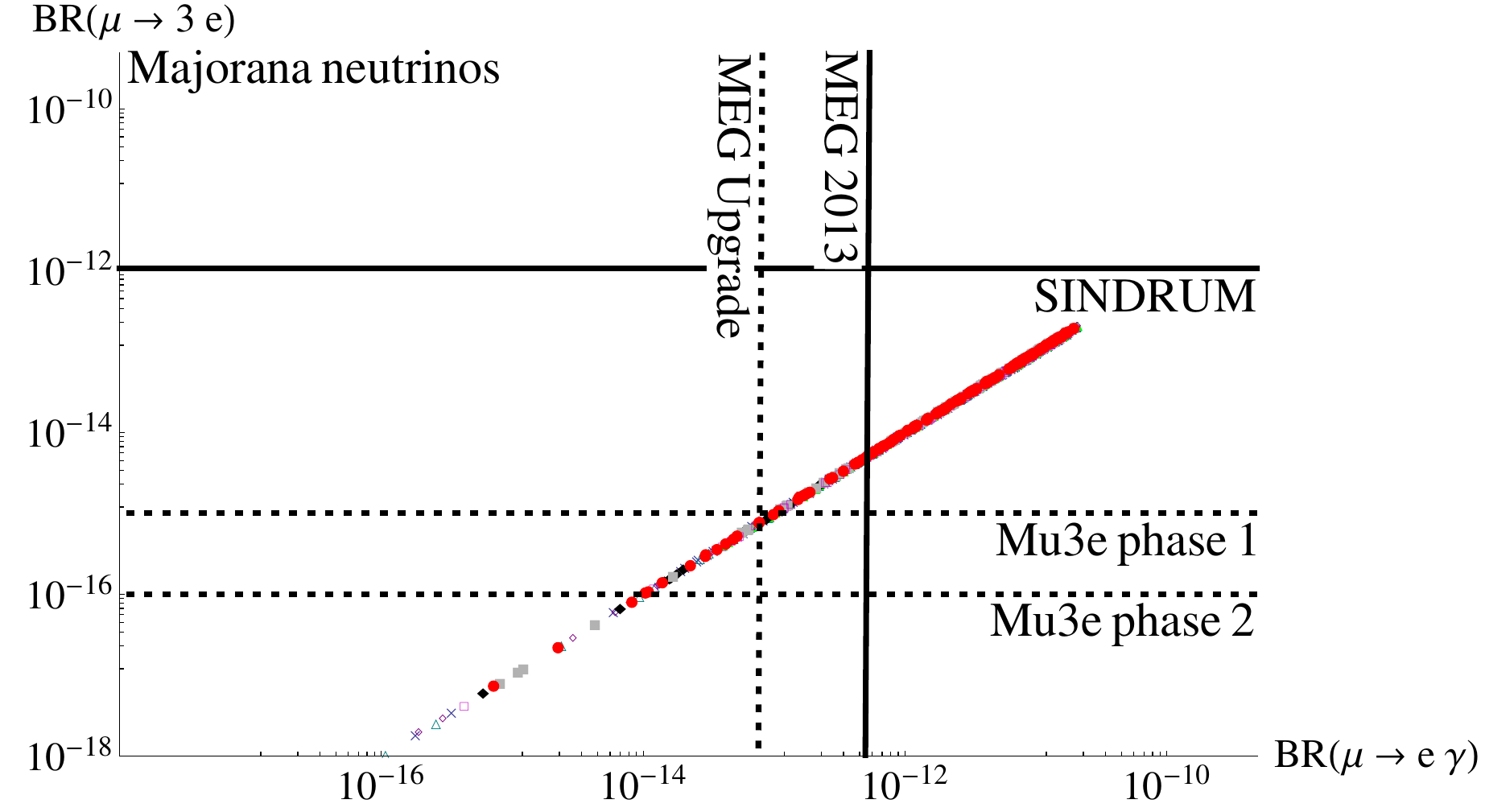}
  \end{center}
  \caption[]{Results for $\mathrm{BR} (\mu\to 3 e)$ versus $\mathrm{BR} (\mu\to e\gamma)$ for the case of Majorana neutrinos. The current experimental bounds from SINDRUM and MEG 2013 are displayed together with the expected future limits
  from Mu3e phase 1 (2) and MEG Upgrade, respectively. For further details see caption of figure \ref{fig3}.}
 \label{fig4}
  \end{figure}

    In the model with Majorana neutrinos lepton flavour violating processes are of the same order as in the SM with neutrino masses and thus completely negligible, since all flavour violation is encoded in the 
   UV-localized mass term for RH neutrinos. This is different in the model for Dirac neutrinos, because the presence of lightish Kaluza-Klein states induces sizable flavour violation at one-loop. A (further) source
   of sizable flavour violation in both models are boundary kinetic terms. The dominant one is $\bar L_L(x,R) (R \hat Z_l) i \, {\slash \!\!\!\! D} L_L(x,R)$ for LH leptons which is localized on the UV brane $z=R$. 
   $\hat Z_l$ is flavour non-diagonal and depends on the chosen $G_f$.
 
\begin{figure}
  \begin{center}
  \includegraphics[width=2.9in]{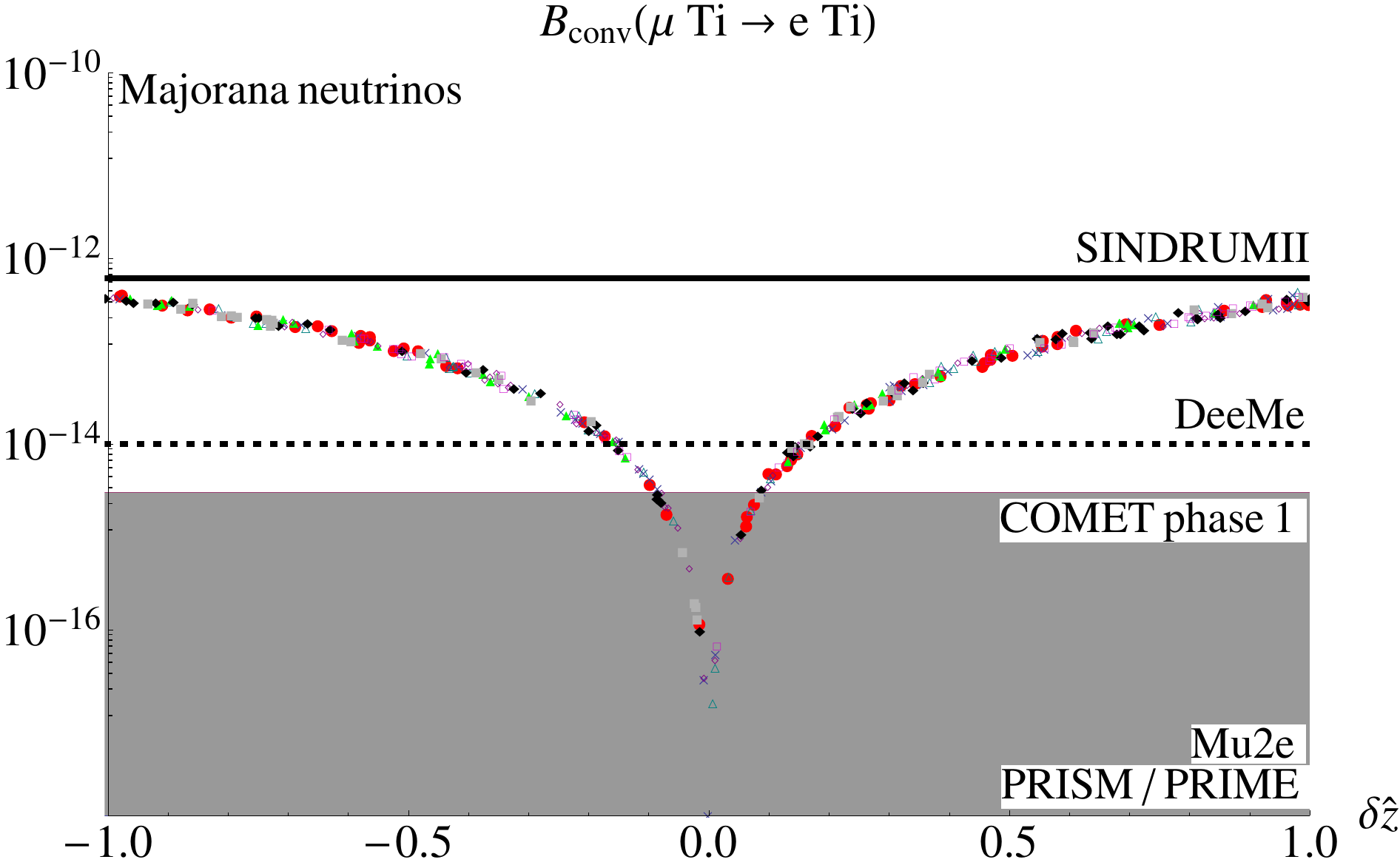}
   \end{center}
   \caption[]{Results for $\mu-e$ conversion in Titanium for the case of Majorana neutrinos for the various choices of the flavour symmetry $X$, see figure \ref{fig3}. The results for Dirac neutrinos are very similar. The current bound is from 
   SINDRUMII and expected future bounds are from DeeMe, COMET phase 1, Mu2e and PRISM/PRIME.}
   \label{fig5}
\end{figure}

  The processes $\mu\to e\gamma$, $\mu\to 3 e$ and $\mu-e$ conversion all depend quadratically on $\delta\hat z \propto (\hat Z_l)_{e\mu}$, see figures \ref{fig3}-\ref{fig5}. The branching ratio of $\mu\to e\gamma$ is about 
   two orders of magnitude larger in the model for Dirac neutrinos than in the one for Majorana neutrinos, see figure \ref{fig3}. The different contributions to $\mathrm{BR} (\mu\to e\gamma)$ and their possible
   cancellation in the model for Dirac neutrinos are clearly visible in figure \ref{fig3}. 
    In the case of Majorana neutrinos experiments searching for $\mu\to 3 e$ can considerably constrain the parameter space, not accessible to MEG (Upgrade), see figure \ref{fig4}. 
  The results for $\mu-e$ conversion in Titanium are very similar in both types of models and the current experimental bound puts a weak constraint on $\delta\hat z$, see figure \ref{fig5}.
   $\mathrm{BR} (\tau\to \mu\gamma)$ and $\mathrm{BR} (\tau\to e\gamma)$ are smaller than $2 \times 10^{-9}$ ($2 \times 10^{-11}$) and $6 \times 10^{-10}$ ($7 \times 10^{-12}$) in the model
   for Dirac (Majorana) neutrinos and thus hardly observable in the near future.

\vspace{0.05in}
{\bf Acknowledgements}: The author would like to thank the organizers of the ``$1^{\mathrm{st}}$ Conference on Charged Lepton Flavor Violation". She is supported by the ERC Advanced 
Grant no. 267985, ``Electroweak Symmetry Breaking, Flavour and Dark Matter: One Solution for Three Mysteries" (DaMeSyFla).



%% References

\nocite{*}
\bibliographystyle{elsarticle-num}
\bibliography{poster}






% \end{thebibliography}

\end{document}